\documentclass[prl,twocolumn,showpacs,superscriptaddress,floatfix]{revtex4}
\usepackage{graphicx}

\begin{document}
\title{Impact of weak localization in the time domain}

\author{S.K.~Cheung}
\affiliation{Department of Physics, Hong Kong University of
Science and Technology, Clear Water Bay, Kowloon, Hong Kong}

\author{X.~Zhang}
\affiliation{Department of Physics, Hong Kong University of
Science and Technology, Clear Water Bay, Kowloon, Hong Kong}

\author{Z.Q.~Zhang}
\affiliation{Department of Physics, Hong Kong University of
Science and Technology, Clear Water Bay, Kowloon, Hong Kong}

\author{A.A.~Chabanov}
\affiliation{Department of Physics, Queens College of the
City University of New York, Flushing, New York 11367, USA}

\author{A.Z.~Genack}
\affiliation{Department of Physics, Queens College of the
City University of New York, Flushing, New York 11367, USA}

\date{\today}

\begin{abstract}
We find a renormalized ``time-dependent diffusion coefficient", $D(t)$,
for pulsed excitation of a nominally diffusive sample by solving the
Bethe-Salpeter equation with recurrent scattering. We observe a crossover
in dynamics in the transformation from a quasi-1D to a slab geometry implemented by
varying the ratio of the radius, $R$, of the cylindrical sample with
reflecting walls and the sample length, $L$. Immediately after the
peak of the transmitted pulse, $D(t)$ falls linearly with a nonuniversal
slope that approaches an asymptotic value for $R/L\gg 1$. The value of
$D(t)$ extrapolated to $t = 0$ depends only upon the dimensionless
conductance $g$ for $R/L\ll 1$ and only upon $k\ell_{0}$ for $R/L\gg 1$,
where $k$ is the wave vector and $\ell_{0}$ is the bare mean free path.
\end{abstract}


\maketitle

Weak localization (WL) of electronic and classical waves arises from
the interference of counterpropagating partial waves in closed loops.
Its impact upon electronic conductance \cite{Mesobook} has been widely
studied using steady-state methods such as magnetoresistance \cite{Bergmann},
and can be directly visualized as enhanced retroreflection of light from
random samples \cite{CBS}. Partial waves with trajectories over a wide
range of lengths contribute to these phenomena. It is, however, of great
interest to investigate the variation of WL upon pathlength
\cite{Altshuler,Berkovits90,Weaver94,Mirlin00,Chabanov02} since the
impact of WL presumably increases with pathlength. Though it has not
been practical to isolate paths of specific lengths in studies of
electronic conductance, this can be accomplished for classical waves
in time-resolved measurements of pulsed transmission through random
media \cite{Chabanov02,McCall87,GenDrake89,Alfano90,Lagendijk97,Zang99}.
Time-resolved transmission measurements \cite{McCall87,GenDrake89}
have generally been consistent with diffusion theory, which predicts
a simple exponential decay of the average transmission following an
initial rise, with the decay rate due to leakage from the sample of
$1/t_{D}=\pi^{2}D_{0}/(L+2z_{0})^{2}$, where $t_{D}$ is the diffusion time,
$D_{0}$ is the diffusion coefficient, $L$ is the sample thickness,
and $z_{0}$ is the extrapolation length.
Recently, however, Chabanov \textit{el al.} \cite{Chabanov02} observed
nonexponential decay of pulsed microwave transmission in a quasi-1D
sample which they characterized by a ``time-dependent diffusion
coefficient", $D(t)$. The decrease in the decay rate with time was
compared to the leading correction term in a supermatrix model
calculation by Mirlin \cite{Mirlin00} of the tail of the electron
survival probability. In the case of the orthogonal ensemble, it can be expressed
in term of the renormalization of a ``time-dependent diffusion coefficient",
$D(t)/D_{0}=1-(2/\pi^{2}gt_{D})t$, where $g$ is the dimensionless conductance.
A linear decay of $D(t)$ was found in microwave experiments; however, $D(t)$ did not
extrapolate to the bare diffusion coefficient at $t=0$ and the scaling behavior of
the slope of $D(t)$ did not agree with the above expression. Nonexponential decay
of pulsed transmission has also been reported recently in numerical
simulations in 2D \cite{Haney03} and in a self-consistent diffusion
theory, which includes recurrent scattering \cite{Bart03}. For times
much larger than the Heisenberg time, $t_H=gt_D$, it has been
predicted that the decay of the electron survival probability follows a
log-normal behavior due to the presence of prelocalized states
\cite{Altshuler}. These long-lived states might be associated with rare
configurations of disorder in the medium \cite{Apalkov02}.

In this Letter, we solve the Bethe-Salpeter equation with recurrent scattering included in a self-consistent manner that satisfies the Ward
Identity \cite{Kirkpatrick85} to obtain the average time-dependent intensity transmitted through a random sample following pulsed
excitation. The sample is cylindrical with reflecting walls. By changing the ratio of the longitudinal dimension $L$ and the radius $R$, a
smooth transition can be made between two key experimental geometries: a quasi-1D geometry with $L\gg R$, which is commonly employed in
microwave experiments, and a slab with $L\ll R$, which is the typical optical geometry. For the quasi-1D geometry, we find $D(t)/D_{0} = A
-(2B/\pi^2gt_{D})t$ for $t\ll t_H$. The constant part, $A$, is universal, depending only on $g$, while $B$ is nonuniversal and depends upon
$L/\ell_{0}$, where $\ell_{0}$ is the bare mean free path, as well as upon $g$. In the limit $g\gg 1$ and $L/\ell_{0}\gg 1$, our results
coincide with supersymmetry calculations by Mirlin \cite{Mirlin00}. For intermediate values of $g$, $g\ge 5$, our results are in agreement
with experiment \cite{Chabanov02}. For the slab geometry, $D$ approaches a nearly constant renormalized value, being equal to
$D_{0}(1-1.03/(k\ell_{0})^2)$ in the limit $L/\ell_{0}\gg 1$. This is close to the result of WL theory for a bulk system.

We consider a scalar wave incident on the front surface of the
random sample at $z=0$. We assume that the medium possesses neither
absorption nor gain and that scattering is isotropic. The time
evolution of intensity within the sample is obtained from the
Fourier transform of the frequency correlation function
$C_{\Omega}(\omega,\mathbf{r})=\langle\phi_{\Omega^{+}}(\mathbf{r}) \,
\phi^{*}_{\Omega^{-}}(\mathbf{r})\rangle$, where
$\Omega^{\pm}=\Omega\pm\omega/2$, $\Omega$ is the wave frequency,
$\omega$ is the frequency shift, and $\phi_{\Omega}(\mathbf{r})$ is
the wavefunction or field inside the sample \cite{Zang99, GenDrake89}.
The function $C_{\Omega}(\omega,\mathbf{r})$ is obtained from the
space-frequency correlation function
$C_{\Omega}(\omega,\mathbf{r},\mathbf{r}')=\langle
\phi_{\Omega^{+}}(\mathbf{r}) \, \phi^{*}_{\Omega^{-}}(\mathbf{r}')\rangle$,
which satisfies the Bethe-Salpeter equation,

\begin{eqnarray}
C_{\Omega}(\omega ,\mathbf{r},\mathbf{r}')\!\!\!
&=&\!\!\!\langle\phi_{\Omega^{+}}(\mathbf{r})\rangle \,
\langle\phi^{*}_{\Omega^{-}}(\mathbf{r}')\rangle \nonumber \\
+ \int\!\!\!\! &d\mathbf{r}_{1}&\!\!\! d\mathbf{r}_{2} \,
d\mathbf{r}_{3} \, d\mathbf{r}_{4} \langle G_{\Omega^{+}}
(\mathbf{r},\mathbf{r}_{1})\rangle\langle \,
G_{\Omega^{-}}(\mathbf{r}',\mathbf{r}_{3})\rangle \, \nonumber \\
&\times& \!\!\! U_{\Omega}(\omega \, ;\mathbf{r}_{1},\mathbf{r}_{2}
\, ;\mathbf{r}_{3},\mathbf{r}_{4}) \, C_{\Omega}(\omega
,\mathbf{r}_{2},\mathbf{r}_{4}) \, , \label{}
\end{eqnarray}
where $\langle\phi_{\Omega}(\mathbf{r})\rangle$ represents the
coherent source inside the sample and
$\langle G_{\Omega}(\mathbf{r},
\mathbf{r}_{1}) \rangle = -{\exp(iK|\mathbf{r}-\mathbf{r}_{1}|)\over4\pi
|\mathbf{r}-\mathbf{r}_{1}|}$ is the ensemble-averaged Green's function.
Here $K=k+{i\over 2\ell}$, where $k={\Omega\over v}$, $v$ is the wave
speed, $\ell$ is the scattering mean free path determined from the
imaginary part of the self-energy of $\langle G\rangle$. $\ell_{0}$
is determined from the single-scattering diagram only. The vertex function
$U_{\Omega}$ in Eq.~(1) represents the sum of all irreducible vertices.
Here we approximate $U_{\Omega}$ as
\begin{eqnarray}
\lefteqn{U_{\Omega}(\omega;\mathbf{r}_{1},\mathbf{r}_{2};
\mathbf{r}_{3},\mathbf{r}_{4})=}\nonumber\\
& &{4\pi\over\ell_{0}}[1+\delta(\omega,k)] \,
\delta(\mathbf{r}_{1}-\mathbf{r}_{2}) \,
\delta(\mathbf{r}_{1}-\mathbf{r}_{3}) \,
\delta(\mathbf{r}_{1}-\mathbf{r}_{4}).
\label{}
\end{eqnarray}
The first term, proportional to $4\pi/\ell_{0}$, generates all self-avoiding
multiple-scattering diagrams, which produce the diffusion result
when $L\gg\ell_{0}$ \cite{Lagendijk88}, whereas the factor
$\delta(\omega,k)$ in the second term represents the WL
contribution to $U_{\Omega}$. The presence of the second
term in the vertex function renormalizes the mean free path,
giving $\ell(\omega,k)=\ell_{0}/[1+\delta(\omega,k)]$. The
Ward Identity, which enforces flux conservation, requires
that the same $\ell(\omega,k)$ should appear in $\langle G\rangle$
of Eq.~(1). Here, we assume
that the system is far from the localization
threshold and that the renormalized mean free path is scale
independent. In a bulk system, $\delta(\omega,k)$ can be
obtained from the renormalized frequency-dependent
diffusion coefficient \cite{Kirkpatrick85},
\begin{equation}
\frac{1}{D(\omega,k)}={\frac{1}{D_0}}
[1+\frac{2\pi v}{k^2}\tilde{G}(\mathbf{r},\mathbf{r};\omega)],
\label{}
\end{equation}
where $D(\omega,k)=v\ell(\omega,k)/3$, and $\tilde{G}$ is
the Green's function for the diffusion equation and reflects
the return probability. Here, we solve for $\tilde{G}$ by using
the boundary conditions of a reflecting tube. After
taking the spatial average, we find
\begin{equation}
\delta(\omega,k)=\delta_{1}(\omega,k)+\delta_{2}(\omega,k),
\end{equation}
with
\begin{equation}
\delta_1(\omega,k)=\frac{v}{2N\tilde{L}}
\sum^{n_c}_{n=1}\frac{1}{-i\omega+D_{0}q_n^2}
\label{}
\end{equation}
and
\begin{equation}
\delta_2(\omega,k)=\frac{3}{2k^2l_0\tilde{L}}\sum^{n_c}_{n=1}
\ln\!\left(\frac{-i\omega+D_{0}(q_n^2+\alpha^2/l_0^2)}
{-i\omega+D_{0}(q_n^2+\pi^2/R^2)}\right),
\label{}
\end{equation}
where $N=(kR)^2/4$ is the number of transverse modes, $q_n=n\pi/\tilde{L}$, $n_c=\alpha\tilde{L}/\pi l_0$, and $\tilde{L}=L+2z_0$ with
$z_0=0.71\ell_{0}$ \cite{Morse}. In the above equations, $\alpha/\ell_{0}$ denotes the upper momentum cutoff, $q_{c}$, and is chosen as
$1/\ell_{0}$ in our calculations unless specified otherwise. The $\delta_{1}$ term arises from diffusive modes which are uniform in the
transverse directions, i.e., $\vec{q}_{\perp}=0$. When $L/\ell_{0}$ is large, the static limit of this term becomes
$\delta_{1}(\omega=0)=1/3g$, where $g=4N\ell_{0}/3\tilde{L}$. This term is responsible for the linear decrease of the decay rate in the
transmitted intensity $\langle I(t)\rangle$. The $\delta_{2}$ term represents the contribution from other transverse modes below
$\vec{q}_{\perp,c}$. This term becomes important only when $R\gg\ell_{0}$ and is responsible for the renormalization of diffusion constant
in the limit of a slab geometry. In our calculation, we replace $D_{0}$ in Eqs.~(5) and (6) by $v\ell(\omega,k)/3$ and solve for
$\delta(\omega,k)$ in a self-consistent manner.

For simplicity of calculation, we assume both the excitation
intensity and the scattered intensity are uniform in the
transverse cross-section of the tube. Eq.~(1) can now be
written as \cite{Zang99}
\begin{eqnarray}
C_{\Omega}(\omega\!\!\!\!\!&,&\!\!\!\!z)=
\exp\!\left({i\omega z\over v}-{z\over\ell_{0}}\right) \nonumber \\
&+&\!\!{1+\delta(k,\omega)\over 4\pi\ell_{0}}\int_{0}^{L}\!\!dz'
H(\omega ,z-z')\, C_{\Omega}(\omega,z') \, ,
\end{eqnarray}
where
\begin{eqnarray}
H(\omega ,z-z')= \pi\!\int\!\! d\rho^{2}{\exp \!\left[\left({i\omega\over v}
-{[1+\delta(\omega,k)]\over\ell_{0}}\right)\!\sqrt{\rho^{2}+(z-z')^{2}}\right]
\over \rho^{2}+(z-z')^{2}} \, .
\end{eqnarray}
Eq.~(7) is solved numerically for $C_{\Omega}(\omega, L)$.
Its Fourier transform gives $\langle I(t)\rangle$, from which
we can calculate the renormalized ``time-dependent diffusion
coefficient", $D(t)/D_{0}=-t_{D}\, d\ln\langle I(t)\rangle/dt$,
for $t>t_{D}$ \cite{Chabanov02}.

We first consider a quasi-1D geometry with $L\gg R\simeq \ell_{0}$. In this case, since $\delta_2$ is nonzero only when
$R>\pi\ell_{0}/\alpha$, we can set $\delta_{2}=0$. For the case of $N=800$, $D(t)/D_{0}$ is plotted in Fig.~1 as a function of the
dimensionless time, $t^{\prime\prime}\equiv 2t/\pi^2gt_{D}$, for $L/\ell_{0}$=20, 40, and 80. Also plotted in Fig.~1 (dotted line) is
Mirlin's analytical result \cite{Mirlin00}, $D(t^{\prime\prime})/D_{0}=1-t^{\prime\prime}$. The decay rate of $D(t^{\prime\prime})/D_{0}$
appears to decrease with increasing $L/\ell_{0}$ and is always smaller than the analytical result. If we fit the linear region to
$D(t^{\prime\prime})/D_{0}=A-Bt^{\prime\prime}$, the intercept $A$ is found to decrease with increasing $L/\ell_{0}$. These observations are
consistent with recent microwave experiments \cite{Chabanov02}. In order to understand the behavior of $A$ and $B$, we have carried out a
systematic study by varying $L/\ell_{0}$ and $N$. The results for $A$ are shown in Fig.~2, where $A$ is plotted as a function of $g$ for
$L/\ell_{0}$=10, 20, and 40. All the data points can be well fit by a single curve, $A=1.00-0.27/g-0.17/g^{2}$, indicating that $A$ is a
function only of $g$. In the limit of large $g$, we recover Mirlin's result, $A=1$. It is interesting to note that the above relation gives,
$\beta=d\ln\! A/d\ln\!\tilde{L}=-(0.27/g+0.34/g^{2})/A$, which corresponds to the exponent of the local scaling relation,
$A\propto\tilde{L}^\beta$. If we use the value in the microwave experiments \cite{Chabanov02} of $g=5.5$, we find $\beta=-0.06$, which is in
good agreement with the measured value of $-0.05$.

\begin{figure}
\includegraphics [width=\columnwidth] {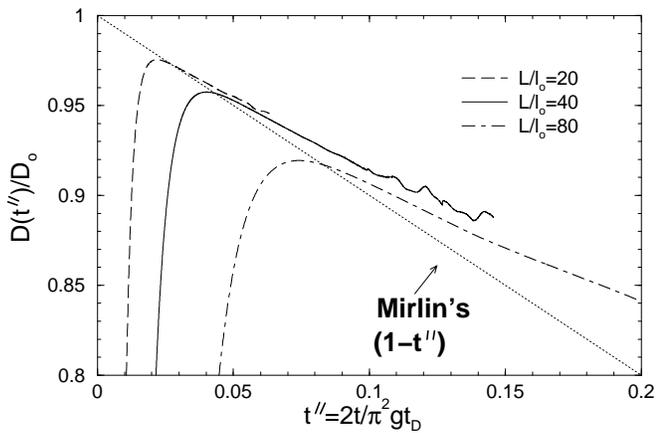}
\caption{$D(t^{\prime\prime})/D_{0}$ plotted as a function of the
dimensionless time, $t^{\prime\prime}\equiv 2t/\pi^2gt_D$, for
quasi-1D samples with $N=800$. The dotted line is the analytical
result of Ref.~[7], $1-t^{\prime\prime}$.}
\end{figure}
\begin{figure}
\includegraphics [width=\columnwidth] {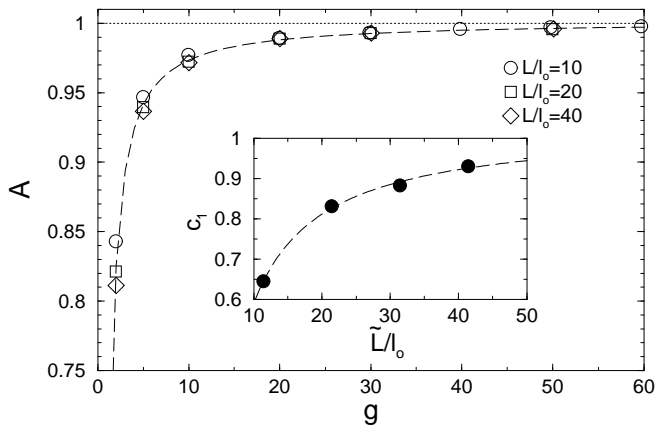}
\caption{The parameter $A$ obtained from the fit of $D(t^{\prime\prime})/D_{0}$ with $A-Bt^{\prime\prime}$ is plotted as a function of $g$
for quasi-1D samples. The dashed curve is $A=1.00-0.27/g-0.17/g^{2}$. In the inset: $c_{1}$ versus $\tilde{L}/\ell_{0}$; the dashed curve is
$c_{1}=1.03-4.42\ell_{0}/\tilde{L}$.}
\end{figure}
\begin{figure}
\includegraphics [width=\columnwidth] {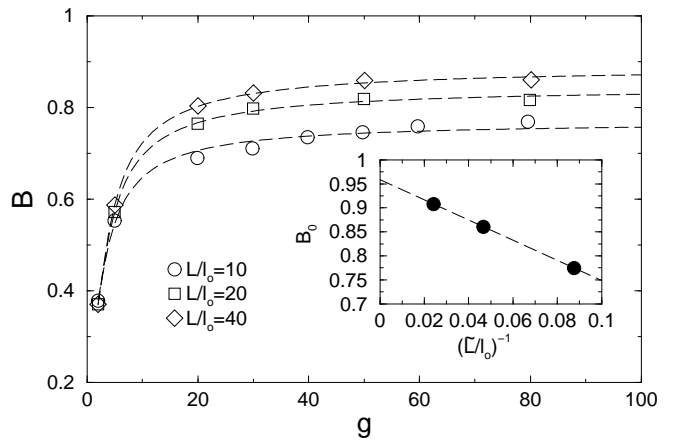}
\caption{The parameter $B$ obtained from the fit of $D(t^{\prime\prime})/D_{0}$ with $A-Bt^{\prime\prime}$ is plotted as a function of $g$
for quasi-1D samples. The dashed curves are the fit of $B$ with $B=B_{0}+B_{1}/g+B_{2}/g^2$, where $B_{i}$ are constant. In the inset:
$B_{0}$ versus $(L/\ell_{0})^{-1}$; the dashed line represents a linear fit.}
\end{figure}
\begin{figure}
\includegraphics [width=\columnwidth] {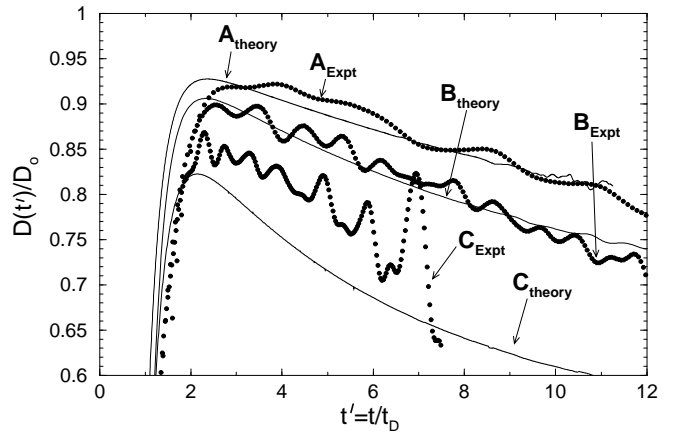}
\caption{$D(t^{\prime\prime})/D_{0}$ plotted as a function of
$t'=t/t_D$ for three samples studied in Ref.~[8], with $N=68$,
$L/\ell_{0}=6.4$ (Sample A), 9.5 (Sample B), and 19.2 (Sample C),
and $D_0=41.5$ cm$^2$/ns. The theoretical results for the
corresponding samples are shown as solid curves.}
\end{figure}
\begin{figure}
\includegraphics [width=\columnwidth] {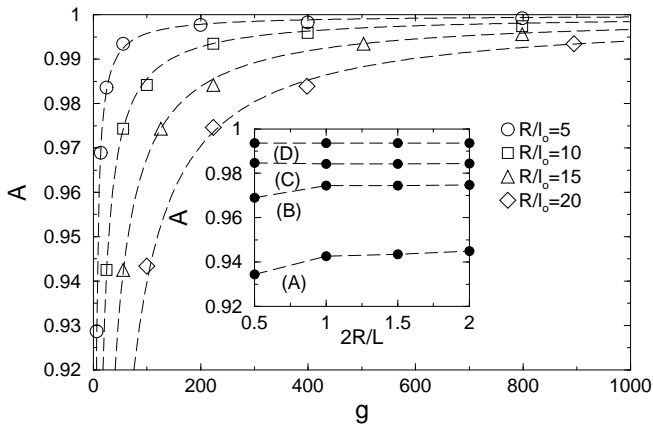}
\caption{The parameter $A$ obtained from the fit of $D(t^{\prime\prime})/D_{0}$ with $A-Bt^{\prime\prime}$ is plotted as a function of $g$
for $L/\ell_0=20$ and different radii $R/\ell_0$. The dashed curves are the fit of $A$ with $A=A_{0}+A_{1}/g+A_{2}/g^2$, where $A_{i}$ are
constant. In the inset: $A$ versus $2R/L$ for $k\ell_0=4$ (A), 6 (B), 8 (C), and 12 (D).}
\end{figure}

The corresponding results for $B$ are shown in Fig.~3. Unlike $A$, $B$ depends not only on $g$ but also on $L/\ell_{0}$. For each
$L/\ell_{0}$, we fit $B$ with $B_{0}+B_{1}/g+B_{2}/g^{2}$, where $B_{i}=B_{i}(L/\ell_{0})$. The values of $B_{0}$ are found to be linear in
$(\tilde{L}/\ell_{0})^{-1}$, as shown in the inset. In the limit of large $L/\ell_{0}$, $B_{0}$ approaches the Mirlin's result of $B=1$.
Thus, our calculations indicate that Mirlin's result is valid only when both $g$ and $L/\ell_{0}$ are sufficiently large. In Fig.~4, we
compare our theory with microwave measurements \cite{Chabanov02} for $N=68$ and $L/\ell_{0}=6.4$ (Sample A), $9.5$ (Sample B), and $19.2$
(Sample C), corresponding to $\ell_{0}=9.5$ cm. The best fit is found for $\alpha=0.5$, and good agreement is found for Samples A and B. For
Sample C, for which $g=3$, however, our theory predicts a smaller decay rate, reflecting the important role of localization effects, which
are not included here.

We next consider the crossover from a quasi-1D to a slab geometry, as $R/L$ increases. When $2R/L$ is sufficiently large, the $\delta_{1}$
term falls leading to a slope of $D(t)$ with a magnitude which decreases as $1/g \propto 1/R^2$. In the limit of large $R/L$, the function
$D(t)$ approaches a nearly constant value, $\tilde{D}$, which is determined by the $\delta_{2}$ term. Thus, we focus our discussion on $A$.
In Fig.~5, we present $A$ as a function of $g$ for different values of $R/\ell_{0}$ for $L/\ell_{0}=20$. These results suggest,
$A=1-c_{0}(L,R)/g$. From the fit (dashed lines) to the data, we find that $c_{0}(L,R)$ is proportional to $R^2$ when $2R>L$, suggesting
$A=1-c_{1}(L)/(k\ell_{0})^2$. This can also be seen in the inset of Fig.~5, where $A$ is replotted as a function of $2R/L$ for different
values of $k\ell_{0}$. It is seen that $A$ approaches a constant value when $2R>L$ for each value of $k\ell_{0}$. In order to find the
behavior of $c_{1}(L)$, we repeat the calculation for $L/\ell_{0}=10$, 30 and 40. The results are plotted in the inset of Fig.~2. The
fitting of these data suggests, $c_{1}=1.03-4.42\ell_{0}/\tilde{L}$ (dashed line), which is the behavior of the renormalized diffusion
coefficient $\tilde{D}$ in a slab. Thus $\tilde{D}$ decreases with increasing $L$, which is due to the presence of longer recurrent
scattering paths in a thicker sample. This behavior is different from that found in thin samples in optical measurements when $L/\ell_{0}<6$
\cite{Lagendijk97}. It is interesting to note that when $L/\ell_{0}\gg 1$, our result $\tilde{D}/D_{0}\simeq 1-1.03/(k\ell_{0})^2$ is close
to the static result of Eq.~(3) in the bulk for $\alpha=1$, i.e., $1-3\alpha/\pi (k\ell_{0})^2$.

In conclusion, we have solved the Bethe-Salpeter equation with recurrent scattering to obtain WL in the time domain. Following peak of the
transmitted pulse, the ``time-dependent diffusion coefficient" $D(t)$ falls nearly linearly. We find the extrapolated value of $D(t)$ at
$t=0$ and the slope of $D(t)$ for samples with different scattering strengths, as the aspect ratio of the sample is changed, transforming
the sample geometry from a quasi-1D to a slab. From this prospective, WL in steady-state measurements can be understood in terms of the
increasing impact of WL associated with trajectories of increasing length.

Discussions with P.~Sheng are gratefully acknowledged.
This research is supported by Hong Kong RGC Grant
No.~HKUST 6163/01P, NSF Grant No.~DMR0205186, and by
U.S. ARO Grant No.~DAAD190010362.

\end{document}